\documentclass[mathleft 
]{an}
\usepackage{graphicx}
\usepackage{times}
\usepackage[sort]{natbib}
\bibpunct{(}{)}{;}{a}{}{,}
\newcommand{\vsini} {$v$\,sin\,$i$}
\newcommand{\kms} {km\,s$^{-1}$}

\newcommand{\ioni}[2]{{#1\,\sc{#2}}}
\newcommand{\macro} {{\em macroturbulence}}
\newcommand{\Teff} {T$_{\rm eff}$}

\begin{document}

\Pagespan{789}{}
\Yearpublication{2006}%
\Yearsubmission{2005}%
\Month{11}%
\Volume{999}%
\Issue{88}%

\title{Is {\em macroturbulent} broadening in OB Supergiants related to pulsations?\,\thanks{Based on 
observations made with the Nordic Optical Telescope, operated
on the island of La Palma jointly by Denmark, Finland, Iceland,
Norway, and Sweden, in the Spanish Observatorio del Roque de los
Muchachos of the Instituto de Astrofisica de Canarias.}}

\author{S. Sim\'on-D\'iaz\inst{1,2}\fnmsep\thanks{Corresponding author:
  \email{ssimon@iac.es}\newline}, K. Uytterhoeven\inst{3}, A. Herrero\inst{1,2}, N. Castro\inst{1,2}, 
        J. Puls\inst{4}
\and  C. Aerts\inst{5,6}
}
\titlerunning{{\em Macroturbulent} broadening in OB Supergiants}
\authorrunning{S. Sim\'on-D\'iaz et al.}
\institute{
Instituto de Astrof\'isica de Canarias, E38200 La Laguna, Tenerife, Spain.
\and 
Departamento de Astrofísica, Universidad de La Laguna, E-38205 La Laguna, Tenerife, Spain.
\and 
Laboratoire AIM, CEA/DSM-CNRS-Universit\'e Paris Diderot; 
CEA, IRFU, SAp, centre de Saclay, 91191, Gif-sur-Yvette, France.
\and 
Universit\"atssternwarte M\"unchen, Scheinerstr. 1, 81679 M\"unchen, Germany
\and 
Instituut voor Sterrenkunde, Katholieke Universiteit Leuven, Celestijnenlaan 200D, 3001 Leuven, Belgium
\and
IMAPP, Department of Astrophysics, Radboud University Nijmegen, PO Box 9010, 6500 GL Nijmegen, the Netherlands}

\received{30 May 2005}
\accepted{11 Nov 2005}
\publonline{later}

\keywords{stars: early-type --- stars: atmospheres --- stars: oscillations ---
          Stars: rotation --- stars: supergiants}

\abstract{%
The spectrum of O and B Supergiants is known to be affected by an important extra 
line-broadening (usually called \macro) that adds to stellar rotation. 
Recent analysis of high resolution spectra has shown that the interpretation
of this line-broadening as a consequence of large-scale turbulent motions would imply 
highly super-sonic velocity fields, making this scenario quite improbable. Stellar 
oscillations have been proposed as a likely alternative explanation.
We present first encouraging results of an observational project aimed at investigating the 
$macroturbulent$ broadening in O and B Supergiants, and its possible connection 
with spectroscopic variability phenomena and stellar oscillations: a) all the studied B Supergiants show 
line profile variations, quantified by means of the first ($\langle v \rangle$) and third velocity ($\langle v^3 \rangle$) 
moments of the lines, b) there is a strong correlation between the peak-to-peak amplitudes 
of the $\langle v \rangle$ and $\langle v^3 \rangle$ variability and the size of the extra-broadening.}

\maketitle

\section{Introduction}

\begin{table*}[t!]
\caption{List of observed stars, their spectral classification, V magnitude, \vsini, 
{\em macroturbulence}, and peak-to-peak amplitude of the first and third moment
from the \ioni{Si}{iii}\,4567 or \ioni{O}{iii}\,5592 lines. Note: \vsini, $\Theta_{\rm G}$,
and $\Delta<v>$ in \kms; $\Delta<v^3>$ in km$^3$\,s$^{-3}$.
}\label{t1}
\centering
\begin{tabular}{lllccccccccc}
\hline \hline
\noalign{\smallskip}
     &      &           &    & & \multicolumn{2}{c}{\vsini\ (FT)} & & {\em Macrot.} & & \multicolumn{2}{c}{LPVs}\\
\cline{6-7} \cline{9-9} \cline{11-12}
\noalign{\smallskip}
Star & Name & SpT \& LC &  V & &  \scriptsize{Range} & \scriptsize{Median} &  & \scriptsize{$\Theta_{\rm G}$} & & $\Delta \langle v \rangle$ & 
$\Delta \langle v^3 \rangle$ x 10$^5$\\
\hline
\noalign{\smallskip}
\multicolumn{9}{l}{Early B-type Supergiants} \\
\hline
\noalign{\smallskip}
HD\,209975 & 19\,Cep           & O9.5\,Iab    & 5.11 & & 54--61 & 57 & & 65 & & 10.1 & 0.99 \\
HD\,37128  & $\epsilon$\,Ori   & B0\,Ia       & 1.70 & & 46--64 & 55 & & 65 & & 12.3 & 0.79 \\
HD\,38771  & $\kappa$\,Ori     & B0.5\,Ia     & 2.05 & & 46--57 & 51 & & 55 & & 9.4  & 0.49 \\
HD\,2905   & $\kappa$\,Cas     & BC0.7\,Ia    & 4.18 & & 44--59 & 52 & & 60 & & 10.1 & 0.63 \\
HD\,190603 &                   & B1.5\,Ia$^+$ & 5.66 & & 23--47 & 39 & & 40 & & 2.2  & 0.15 \\
HD\,14818  & 10\,Per           & B2\,Ia    0  & 6.27 & & 36--47 & 41 & & 50 & & 7.8  & 0.51 \\
\noalign{\smallskip}
\hline
\noalign{\smallskip}
\multicolumn{9}{l}{Dwarfs (Luminosity class V objects)} \\
\hline
\noalign{\smallskip}
HD\,214680 & 10\,Lac           & O9\,V     &  4.88  & & 17--23 & 18 & & 25   & & 2.1 & 0.05 \\
HD\,37042  & $\theta^2$ Ori\,B & B0.5\,V   &  6.02  & & 32--34 & 33 & & $<$5 & & 0.8 & 0.01 \\
\hline \hline
\end{tabular}
\end{table*}

The presence of an important extra line-broadening (in addition to the rotational 
broadening, and usually called \macro) affecting the spectra of O and B Supergiants 
(Sgs) has been confirmed by several authors since the first studies of line-broadening 
in O and B stars early in the 1950's. It was initially suggested by 
the deficit of narrow lined objects among these type of stars (\citealt{Sle56}; \citealt{Con77}; \citealt{How97}).
The advent of high-quality spectra allowed to 
confirm that the rotational broadening alone was not sufficient to fit the line profiles 
in many objects, and to investigate the possible disentangling of both broadening 
contributions \citep[see e.g.][]{Rya02, Sim07}. 
These studies definitely showed that, while the effect of \macro\ in OB dwarfs 
is usually negligible when compared to rotational broadening, the effect of this 
extra-broadening is clear\-ly present in OB\,Sgs.

Despite it was named \macro\ at some point, the interpretation of this extra-broadening 
as the effect of turbulent motions is quite improbable. The effect is present in 
photospheric lines and affects the whole profile, even the wavelengths close to the 
continuum. Therefore, whatever is producing the extra-broadening has to be deeply 
rooted in the stellar photosphere (and maybe below), in layers in which we do not 
expect any significant velocity field in these stars. If interpreted as 
turbulent motions, \macro\ would represent {\em highly supersonic} velocities in many 
cases (\citealt{Duf06}; \citealt{Lef07}; \citealt{Mar08}; \citealt{Fra10}). This interpretation is incompatible with 
the previous statement. 

One physical mechanism suggested to be the origin of this extra-broadening relates to 
oscillations. Many OB\,Sgs are known to show photometric
and spectroscopic variability.
Based on this, \citet{Luc76} postulated that this variability may be a pulsation
phe\-no\-me\-non, and {\em macroturbulence} may be identified with the surface motions
generated by the superposition of numerous nonradial oscillations.
More recently, \citet{Aer09} computed time-series of line profiles for evolved massive stars 
broadened by rotation and thousands of low amplitude nonradial gravity mode 
oscillations and showed that the resulting profiles could mimic the observed ones. 
Stellar oscillations are therefore a plausible explanation for the extra-broadening
in O and B Sgs, but so far there is no direct evidence confirming their 
presence. 

We present first results of an observational project aimed at 
investigating the extra line-broadening in O and B Sgs and its possible 
connection with spectroscopic variability phenomena and stellar oscillations.

\section{The project}
The observing campaigns for this project began already two years ago. We selected a sample of
$\sim$\,15 bright O and B stars, with the objective of obtaining time-series of high resolution,
high signal-to-noise (SNR) spectra. With these time-series
spectra we plan to (1) investigate and quantify the presence of line profile variations (LPVs);
(2) investigate the origin of the LPVs, considering several possible physical explanations
(not only stellar oscillations, but also wind variability in terms of density and/or velocity); 
(3) whenever possible, perform a seismic-like study, de\-ter\-mi\-ning the
frequencies associated with the LPVs and, subsequently, perform a mode indentification
and seismic mo\-de\-ling; (4) obtain the stellar and wind parameters of the selected stars 
through an spectroscopic analysis using the stellar atmosphere code FASTWIND \citep{Pul05};
(5) characterize the line-broadening in photospheric lines by disentangling
the projected rotational velocity (\vsini) and the extra-broadening; (6) investigate
the temporal behavior of these quantities; (7) look for empirical relations between
the size of the extra-broadening, \vsini, wind-variability, and the LPVs; (8) hopefully,
obtain firm observational evidences about the physical origin of the extra-broadening. 

Regarding point (3) above, this project will require an important amount of observational time 
(as it is usual in asteroseismic studies, e.g. \citealt{Aer04}; \citealt{Han06}; \citealt{Uyt08}; \citealt{Bri09}; \citealt{Por09} ). 
Although we have already obtained time-series spectra during several observing 
campaigns (using FIES, SES, and HERMES spectrographs, attached to the NOT, STELLA and 
MERCATOR telescopes, respectively), the a\-mount and time\,span of 
the collected spectra are still not enough for a proper seismic study. New campaigns are 
plan\-ned to improve this situation. Meanwhile, the analysis of the available data\,sets are 
leading to interesting and motivating results.
Here, we summarize the main results obtained from the analysis of the first campaign, with FIES.

\section{First results from the FIES08 run}
\subsection{Selected targets and observational data\,set}
In 2008 November, we obtained a first set of spectra with FIES@NOT in medium resolution mode (R=46000). 
Du\-ring 4 nights we collected time-series of spectra for six OB Sgs. The sample was complemented 
with two OB dwarf stars, where the extra-broadening tends to be negligible with respect 
to the rotational broadening. The exposure times were chosen such as to reach at least 
a SNR=200 (measured in the range 4500\,--\,4600 \AA). The list of 
observed stars, along with their spectral classification and V magnitude, is presented in 
Table \ref{t1}.

\subsection{Characterization of line-broadening in photospheric lines}

   \begin{figure*}[t!]
   \centering
   \includegraphics[width=5.0cm, angle=90]{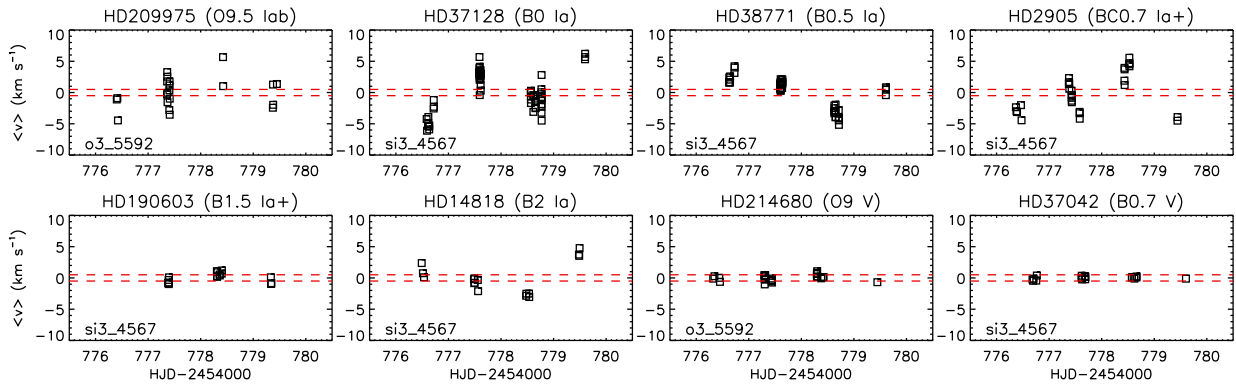}
      \caption{Variation of the first moments (radial velocity placed at average zero) for the sample of stars  
      derived from the \ioni{Si}{iii}\,4567 or \ioni{O}{iii}\,5592 line profiles. Horizontal 
	  lines show the accuracy associated with the instrumental setting used for FIES@NOT.}
         \label{f1}
   \end{figure*}

We used the Fourier transform technique 
(\citeauthor{Gra76} \citeyear{Gra76}; see also \citeauthor{Sim07}, \citeyear{Sim07}
for a recent application to the spectra of O and B stars) to disentangle
the rotational and {\em macroturbulent} broadening contributions. 
The results of the analysis are presented in Table \ref{t1}. We performed
the analysis for each of the time-series spectra, obtaining the \vsini\ indicated
by the first zero of the Fourier transform. The range and median of derived
\vsini\ values are indicated in the first and second columns of Table \ref{t1}. Note that the
dispersion in the obtained \vsini\ is between 10\% and 30\%, depending on the star.
Whether this dispersion is real, or an effect of noise, is not clear from this
data\,set. As outlined by \citet{Sim07}, the correct identification of this 
zero is complicated in the cases of low SNR and a large contribution of 
the extra-broadening. We plan to explore this in more detail in future.

Next, we considered the median \vsini\ va\-lues for each star and quantified the 
contribution of the extra-broadening by assuming a Gaussian-type profile. The 
corresponding va\-lues, indicated as $\Theta_{\rm G}$, are indicated in Table \ref{t1}. 

Note that the extra-broadening is significant in all the Sgs. This is
not the case for the B0.5\,V star HD\,37042, where the total broadening is
mainly produced by the effect of the stellar rotation. The other dwarf star, 
HD\,214680 is a special case, since it has a very low \vsini. For such a low 
\vsini, microturbulence provides a significant contribution to the total 
broadening, which then is included in the measured extra-broadening. 

\subsection{Line profile variations in photospheric lines}

Similarly to previous works studying spectroscopic variabi\-lity in
O and B Sgs (e.g. \citealt{Ebb82}; \citealt{How93}; \citealt{Ful96}; 
\citealt{Pri96}, \citeyear{Pri04}, \citeyear{Pri06}; \citealt{Mor04}; \citealt{Kau06}; \citealt{Mar05}, \citeyear{Mar08}), we
found clear signatures of line profile va\-ria\-tions (LPVs) for all the
Sgs considered in our study.
To quantitatively investigate these LPVs we computed the first, $\langle v \rangle$, and 
third, $\langle v^3 \rangle$, normalised velocity 
moments\footnote{See definition in \citep{Aer10}.}
from the \ioni{Si}{iii}\,4567 or \ioni{O}{iii}\,5592 
lines\footnote{We used FAMIAS \citep{Zim08}, a software package developed 
in the framework of the FP6 European Coordination Action HELAS 
(http://www.helas-eu.org/).}. These moments are connected with the centroid 
velocity of the line and the skewness of the line profile, respectively. 

Results for the first velocity moment\footnote{The third velocity moment 
follows a similar temporal behavior.} are presented in Figure \ref{f1}. 
The associated uncertainties (not included in the plot) are $\sim$\,0.1\,-\,0.4 \kms.
In the case of the dwarf star HD\,37042, the $\langle v \rangle$ values are fairly constant. The maximum 
dispersion in velocity for this stars is $\sim$1 \kms, of the order of the accuracy 
associated with the instrumental setting used for the FIES@NOT observations (indicated 
as red ho\-ri\-zon\-tal lines). All the other stars show $\langle v \rangle$ variations above this significance 
level, with maximum amplitudes $\sim$10--12 \kms. The minimum variations are found for 
HD\,190603 and HD\,214680 (the other luminosity class V object con\-si\-de\-red in this study), with 
maximum amplitudes slightly larger than the significance level.
We indicate in Table \ref{t1} the peak-to-peak amplitude of the first and third moment
variations measured for each of the considered targets.

   \begin{figure}[t!]
   \centering
   \includegraphics[width=4.2cm, angle=90]{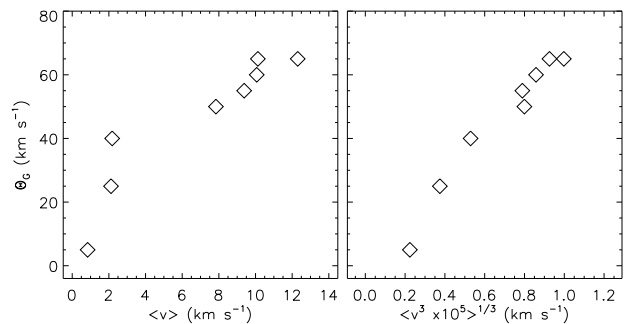}
      \caption{$Macroturbulent$ broadening ($\Theta_{\rm G}$) vs. peak-to-peak amplitude of the first 
	  and third moments of the line profile, for the sample from Table \ref{t1}.}
         \label{f2}
   \end{figure}

\subsection{The {\em macroturbulence}-LPVs connection}

We then investigated the possible connection between the {\em macroturbulent} broadening
and the LPVs. Fig. \ref{f2} shows a clear positive correlation between the size of the 
{\em macroturbulent} broadening ($\Theta_{\rm G}$) and the peak-to-peak amplitude of 
variation of $\langle v \rangle$ and $\langle v^3 \rangle$. 
To our knowledge, this is the {\em first clear observational evidence ever presented for a 
connection between the extra-broadening and the LPVs in OB Sgs.}  
Particular remarkable is the $\Theta_{\rm G}$\,-\,$\langle v^3 \rangle$ correlation: 
the larger the extra-broadening, the more asymmetric line profiles can be found. Note that this
does not mean that lines with an important {\em macroturbulence} contribution are always
asymmetric since $\langle v^3 \rangle$ is oscillating between positive to negative values
with time.

\subsection{Frequency analysis}
We searched for periodic signals in the spectral time-series of the best time-sampled
candidates, and found indications for the presence of at least one long-term period of 0.5 to 3.5 days
in the moment and pixel-to-pixel variations. This allows to discard 
phenomena connected to stellar rotation as the origin of the LPVs (rotational periods for these
stars are of the order of several weeks to a few months). Unfortunately, the 
time span of our observations ($\Delta$T=3.07\,d at best) is not long enough to 
permit a reliable frequency analy\-sis, needed for a subsequent mode identification and
seismic modeling. We hope to improve this situation with future observing runs.

\section{Discussion}
In the last decades, many studies have been performed which aimed at studying and providing empirical 
constraints on the different physical components that can yield temporal variability 
in the photospheric lines of luminous OB stars. It is quite 
common to find in them the suggestion that nonradial oscillations may be 
the origin of the LPVs, and the driver of large-scale wind structures. Observational 
evidence points towards this hypothesis, but a firm confirmation (by means of a 
rigorous seismic analysis) has not been achieved yet.

From a theoretical point of view, g-modes were not initially expected in B\,Sgs because 
the radiative damping in the core was suspected as too strong. \citet{Sai06} 
claimed the detection of simultaneous p- and g-modes in HD\,163899 (B2\,Ib/II) using 
data from the MOST satellite. These authors also computed new models showing that g-modes 
can be excited in massive post-Main sequence stars, as the g-modes are reflected at the 
convective zone asso\-cia\-ted to the H-burning shell. \citet{Lef07} presented observational 
evidence of g-mode instabilities in a sample of photometrically variable B\,Sgs from the 
location of the stars in the (log\,$g$,log\,\Teff)-diagram. A similar conclusion can be 
achieved from the loci of our targets in this diagram.

It seems that macroturbulent broadening in OB Supergiants could be related to pulsations,
but it is too premature to consider them as the only physical phenomenon to explain the
unknown broadening. Nevertheless, our observational study clearly indicates a connection
between {\em macroturbulence} and LPVs, whatever the origin of the latter.\\

\acknowledgements
Financial support by the Spanish Ministerio de Ciencia e Innovaci\'on under the 
project AYA2008-06166-C03-01 and the Consolider-Ingenio 2010 Program grant 
CSD2006 --00070: First Science with the GTC (http://www.iac.es/consolider- ingenio-gtc).
The research leading to these results has also received funding from the European
Research Council under the European Community's Seventh Framework Programme
(FP7/2007--2013)/ERC grant agreement n$^\circ$227224 (PROSPERITY).

\newpage

\end{document}